\documentclass[sigconf]{acmart}

\usepackage[T1]{fontenc}
\usepackage{amsmath}
\usepackage{amsfonts}
\usepackage{mathtools}
\usepackage{stmaryrd}
\usepackage[bb=boondox]{mathalfa}

\usepackage{graphicx}
\usepackage{multicol}
\usepackage{ifdraft}
\usepackage{ifthen}
\usepackage{ebproof}
\usepackage{algorithm}
\usepackage{algorithmicx}
\usepackage{xspace}
\usepackage[noend]{algpseudocode}
\usepackage{tikz, pgf}
\usepackage{hyperref}
\usepackage{url}
\usepackage{subcaption}
\usepackage{lib/latexutils/abbrev}
\usepackage{lib/latexutils/sidenote}
\usepackage{lib/latexutils/symbols}
\usepackage{lib/latexutils/theorems} 

\usepackage[frozencache,cachedir=_minted]{local/minted}

\setminted{
  linenos=true,
  autogobble
}

\ebproofset{
  right label template=\textsc{\inserttext}
}
\ebproofnewstyle{small}{
  separation = 1em, rule margin = .5ex,
  template = \footnotesize$\inserttext$,
  right label template=\footnotesize\textsc{\inserttext}
}

\usetikzlibrary{cd, arrows.meta, decorations.pathmorphing}
\tikzset{
  >=Kite,
  invisible/.style={opacity=0},
  visible on/.style={alt={#1{}{invisible}}},
  alt/.code args={<#1>#2#3}{%
      \alt<#1>{\pgfkeysalso{#2}}{\pgfkeysalso{#3}}%
  }
}
\tikzcdset{
  arrow style=tikz,
  diagrams={>=Kite},
  ampersand replacement=\&,
  labels=description
}

\algrenewcommand\algorithmicindent{1.0em}%
\algrenewcommand\algorithmicfunction{\textbf{fn}}
\algnewcommand\algorithmicyield{\textbf{yield}\ }
\algnewcommand\algorithmiccontinue{\textbf{continue}}
\algnewcommand\algorithmicbreak{\textbf{break}}

\newcommand\Continue{\algorithmiccontinue}
\newcommand\Break{\algorithmicbreak}

\DeclareMathOperator*{\powerset}{\ensuremath{\mathcal{P}}}
\newcommand{\tstr}{\ensuremath{\mathtt{str}}}

\newcommand\omstarr{\ensuremath{\omega^{\star}_r}}

\ifthenelse{\boolean{true}}{%
\newcommand\freetrr[1]{\ensuremath{\xmapsto[r\star]{#1}}}
\renewcommand\freetr[1]{\ensuremath{\xmapsto[a\star]{#1}}}
\newcommand\trr[1]{\ensuremath{\xmapsto[r]{#1}}}
}{%
\newcommand\freetrr[1]{\ensuremath{\xmapsto[\omstarr]{#1}}}
\renewcommand\freetr[1]{\ensuremath{\xmapsto[\omstara]{#1}}}
\newcommand\trr[1]{\ensuremath{\xmapsto[\omega_r]{#1}}}
}

\newcommand\abstractr{\ensuremath{\mathit{abstract}_r}}

\newboolean{preprint}
\newboolean{submission}

\renewcommand\eps{\ensuremath{[]}}

\setboolean{preprint}{true}
\setboolean{submission}{false}

\begin{document}
  \begin{abstract}
  \emph{Constraint Handling Rules} (CHR) is a rule-based programming language which is typically embedded into a general-purpose language.
  There exists a plethora of implementations of CHR for numerous host languages.
  However, the existing implementations often re-invent the way to embed CHR, which impedes maintenance and weakens assertions of correctness.
  To formalize and thereby unify the embedding of CHR into arbitrary host languages, we introduced the framework \emph{FreeCHR} and proved it to be a valid representation of classical CHR.
  Until now, this framework only includes a translation of the \emph{very abstract} operational semantics of CHR which, due to its abstract nature, introduces several practical issues.
  In this paper, we introduce an execution algorithm for FreeCHR.
  We derive it from the \emph{refined} operational semantics of CHR, which resolve the issues introduced by the very abstract semantics.
  We also prove soundness of the algorithm \WRT the very abstract semantics of FreeCHR.
  Hereby we provide a unified and an easy to implement guideline for new CHR implementations, as well as an algorithmic definition of the \emph{refined} operational semantics.
  \ifthenelse{\boolean{preprint}}{This is a preprint of a paper submitted to the \emph{27th International Symposium on Principles and Practice of Declarative Programming}.}
\end{abstract}

  \title{An instance of FreeCHR with refined operational semantics}

\author{Sascha Rechenberger}
\email{sascha.rechenberger@uni-ulm.de}
\orcid{0009-0003-5737-5831}
\affiliation{%
    \institution{Institute for Software Engineering and Programming Languages, Ulm University}
    \city{Ulm}
    \state{Baden-Württemberg}
    \country{Germany}
}

\author{Thom Frühwirth}
\email{thom.fruehwirth@uni-ulm.de}
\affiliation{%
    \institution{Institute for Software Engineering and Programming Languages, Ulm University}
    \city{Ulm}
    \state{Baden-Württemberg}
    \country{Germany}
}

\setcopyright{acmlicensed}
\copyrightyear{2025}
\acmYear{2025}
\acmDOI{XXXXXXX.XXXXXXX}
\acmConference[PPDP 2025]{27th International Symposium on Principles and Practice of Declarative Programming}{September 10-11, 2025 }{University of Calabria, Rende, Italy}
  \maketitle


  \section{Introduction}
\emph{Constraint Handling Rules} (CHR) is a rule-based programming language that is usually embedded into a general-purpose language.
Having a CHR implementation available enables software developers to solve problems in a declarative and elegant manner.
Aside from the obvious task of implementing constraint solvers \cite{fruehwirth2006complete,dekoninck2006inclp}, it has been used, \EG to solve scheduling problems \cite{abdennadher2000university}, and implement concurrent and multi-agent systems \cite{thielscher2002reasoning,thielscher2005flux,lam2006agent,lam2007concurrent}.
In general, CHR is ideally suited for any problem that involves the transformation of collections of data.
Programs consist of rewriting rules which hides away the process of finding suitable candidates for rule application.
Hereby, we get a purely declarative representation of the algorithm without the otherwise necessary boilerplate code.

The literature on CHR as a formalism consists of a rich body of theoretical work, including a rigorous formalization of its declarative and operational semantics \cite{fruehwirth2009constraint,sneyers2010time,fruehwirth2015constraint}, relations to other rule-based formalisms \cite{fruehwirth2025principles} and results on properties like confluence \cite{christiansen2015confluence,gall2017decidable}.

Implementations of CHR exist for a number of languages, such as Prolog \cite{schrijvers2004leuven}, C \cite{wuille2007cchr}, C++ \cite{barichard2024chr}, Haskell \cite{chin2008typesafe,lam2007concurrent}, JavaScript \cite{nogatz2018chr} and Java \cite{abdennadher2002jack,vanweert2005leuven,ivanovic2013implementing,wibiral2022javachr}.

While the implementations adhere to the formally defined operational semantics, they are not direct implementations of a common formal model.
Therefore, the two aspects of CHR (formalism and programming language) are not strictly connected with each other and
there is hence no strict guarantee that the results on the formalism CHR are applicable on the programming language CHR.
Although, such a strict connection is probably not entirely achievable (unless we define and use everything inside a proof assistant like \emph{Coq} or \emph{Agda}), it is desirable to have both formal definition and implementation as closely linked as possible.
In addition to being able to directly benefit from theoretical results, implementors of CHR embeddings and users of the language can also use the formally defined properties to validate their software, \FI in property-based testing frameworks like \emph{QuickCheck}\footnote{\url{https://hackage.haskell.org/package/QuickCheck}} or \emph{jqwik}.\footnote{\url{https://jqwik.net/}}

Another apparent issue within the CHR ecosystem is that many of the implementations of CHR are currently unmaintained.
Although some of them are mere toy implementations, others might have been of practical use.
One example is JCHR \cite{vanweert2005leuven} which would be a useful tool if it was kept on par with the development of Java, especially with modern build tools like \emph{Gradle}.
Having a unified formal model from which every implementation is derived could ease contributing to implementations of CHR as it provides a strict documentation and description of the system, a priori.
Also, different projects might even be merged which would prevent confusion due to multiple competing, yet very similar implementations, as it can be observed in the \emph{miniKanren} ecosystem (\EG there exist about 20 implementations of \emph{miniKanren} dialects for Haskell alone\footnote{https://minikanren.org/\#implementations}).

A third major issue is that many implementations, like the aforementioned JCHR or CCHR \cite{wuille2007cchr}, are implemented via an external embedding, \IE they rely on a separate compiler which translates CHR code into code of the host language.
Although modern build-tools like \emph{Gradle} simplify the inclusion of external tools, every new link in the build-chain is still a nuisance and an additional point of failure.
In contrast, an internal embedding, \IE an embedding of the language via constructs provided by the host language, can easily be implemented as a library.
Such a library can be distributed via a package repository (which exist for most modern programming languages) and handled as a dependency by the respective build-tool.
This dramatically simplifies the use of an embedded language, compared to an external embedding.
Examples of this are the \emph{K.U. Leuven CHR system}, which is implemented as a library in Prolog and distributed as a standard package with \emph{SWI-Prolog}\footnote{\url{https://www.swi-prolog.org/pldoc/man?section=chr}}, or by the library \emph{core.logic} which implements \emph{miniKanren} for the LISP dialect \emph{Clojure}.\footnote{\url{https://github.com/clojure/core.logic}}

The framework \emph{FreeCHR} was introduced to solve the issues discussed above \cite{rechenberger2023freechr}.
It formalizes the embedding of CHR via \emph{initial algebra semantics}.
This common concept in functional programming is used to inductively define languages and their semantics \cite{hudak1998modular,johann2007initial}.
FreeCHR provides both a guideline and high-level architecture to implement and maintain CHR implementations across host languages.
It also creates a strong connection between the practical and formal aspects of CHR.
Until now the framework only formalizes the \emph{very abstract} operational semantics of CHR and lags behind in terms of practical expressiveness.
Actual implementations of CHR typically implement the \emph{refined} operational semantics which were first formalized by Duck et al. \cite{duck2004refined}.
The \emph{refined} operational semantics resolve most sources of non-determinism like the order in which the program and its rules are traversed for matching values.
This makes programs more controllable and hence allows more optimizing programming techniques.

In this paper, we introduce an execution algorithm for FreeCHR which we will derive from the \emph{refined} operational semantics of CHR.
We will also show soundness \WRT the \emph{very abstract} operational semantics of FreeCHR to show that the algorithm constitutes a valid concretization.

The presented algorithm will serve a twofold purpose:
\begin{itemize}
    \item It provides an easy to implement baseline of practically useable FreeCHR implementations,
    \item and an algorithmic definition of the \emph{refined} operational semantics for FreeCHR for formal considerations.
\end{itemize}
To our knowledge, the presented execution algorithm and its implementations will also be the first of the full language definition of ground CHR for which there are formal proofs of correctness.

The rest of the paper is structured as follows:
\autoref{sec:functors} introduces necessary preliminary definitions and notations,
\autoref{sec:non-herbrand-chr} introduces the \emph{refined} operational semantics of CHR
and \autoref{sec:freechr} introduces FreeCHR and accompanying definitions.
The reminder of the paper contains our new contribution: \autoref{sec:ExtendedStates} defines the necessary data structures and \autoref{sec:instance} the FreeCHR execution algorithm with \emph{refined} operational semantics.
\autoref{sec:soundness} provides the central theorem of this paper stating soundness \WRT the \emph{very abstract} operational semantics.
\autoref{sec:related-work} discusses related work,
\autoref{sec:limitations} limitations of our approach and
\autoref{sec:future_work} future work.
Finally, \autoref{sec:conclusion} concludes the paper.

  \section{Preliminaries}\label{sec:functors}
In this section, we introduce preliminary concepts from category theory which we will introduce as instances in the category of sets \textbf{Set}.
We will also introduce our notations for labelled transition systems.

\subsection{Basic definitions}
The \emph{disjoint union} of two sets $A$ and $B$
\begin{align*}
  A \sqcup B = \left\{l_A(a) \mid a \in A\right\} \cup \left\{l_B(b) \mid b \in B\right\}
\end{align*}
is the union of both sets, with additional labels $l_A$ and $l_B$ added to the elements to keep track of the origin set of each element.
We will also use the labels $l_A$ and $l_B$ as \emph{injection} functions $l_A : A \rightarrow A \sqcup B$ and $l_B : B \rightarrow A \sqcup B$ which construct elements of $A \sqcup B$ from elements of $A$ or $B$, respectively\footnote{We will omit labels if their origin set is clear from the context.}.

For two functions $f : A \rightarrow C$ and $g : B \rightarrow C$, the function
\begin{align*}
  \left[f,g\right] &: A \sqcup B \rightarrow C\\
  \left[f,g\right](l(x)) &=\left\{\begin{matrix*}[l]
    f(x), & & \mbox{if}\ l = l_A \\
    g(x), & & \mbox{if}\ l = l_B \\
  \end{matrix*}\right.
\end{align*}
is called a \emph{case analysis} function of the disjoint union $A \sqcup B$.
It is a formal analogue to a \texttt{case-of} expression.
Furthermore, we define two functions
\begin{align*}
  f \sqcup g &: A \sqcup B \rightarrow A' \sqcup B'\\
  (f \sqcup g)(l(x)) &= \left\{
    \begin{matrix*}[l]
      l_{A'}(f(x)), & &\mbox{if}\ l = l_A \\
      l_{B'}(g(x)), & &\mbox{if}\ l = l_B
    \end{matrix*}
  \right.
  \\
  \\
  f \times g &: A \times B \rightarrow A' \times B'\\
  (f \times g)(x,y) &= (f(x), g(y))
\end{align*}
which lift two functions $f : A \rightarrow A'$ and $g : B \rightarrow B'$ to the disjoint union and the Cartesian product, respectively.
We use these concepts to construct abstract union and product types.

\subsection{Endofunctors and $F$-algebras}
A \textbf{Set}-endofunctor\footnote{Since we only deal with endofunctors in Set, we will simply call them \emph{functors}.} $F$ maps all sets $A$ to sets $F A$ and all functions $f : A \rightarrow B$ to functions $F f : F A \rightarrow F B$, 
such that
\begin{align*}
  F \id_A = \id_{F A}
\end{align*}
where $F (g \circ f) = F g \circ F f$.
$\id_X(x) = x$ is the identity function on a set $X$\footnote{We will omit the index of $\id$, if it is clear from the context.}.
A signature $\Sigma = \left\{\sigma_1/a_1,...,\sigma_n/a_n\right\}$, where $\sigma_i$ are operators and $a_i$ their arity, generates a functor
\begin{align*}
  F_{\Sigma} X = \bigsqcup_{\sigma/a\in\Sigma} X^{a} && 
  F_{\Sigma} f = \bigsqcup_{\sigma/a\in\Sigma} f^{a}
\end{align*}
$X^{n}$ and $f^{n}$ are defined as
\begin{align*}
  X^{n} = \underbrace{X \times ... \times X}_{\text{$n$ times}} && f^{n} = \underbrace{f \times ... \times f}_{\text{$n$ times}}
\end{align*}
with $X^0 = \mathbb{1}$ and $f^0 = \id_{\mathbb{1}}$.
$\mathbb{1}$ is a singleton set.
Such a functor $F_{\Sigma}$ models \emph{flat} (\IE not nested) terms over the signature $\Sigma$.
We will use endofunctors to abstractly model the syntax of FreeCHR later in \autoref{sec:freechr}.

Since an endofunctor $F$ defines the syntax of terms, an evaluation function $\alpha : F A \rightarrow A$ defines the \emph{semantics} of terms.
We call such a function $\alpha$, together with its \emph{carrier} $A$, an $F$-algebra $(A, \alpha)$.

If there are two $F$-algebras $\left(A, \alpha\right)$ and $\left(B, \beta\right)$ and a function
\begin{align*}
  h : A \rightarrow B
\end{align*}
we call $h$ an \emph{$F$-algebra homomorphism}, \IFF
\begin{align*}
  h \circ \alpha = \beta \circ F h
\end{align*}
\IE $h$ preserves the structure of $\left(A, \alpha\right)$ in $\left(B, \beta\right)$ when mapping $A$ to $B$.
In this case, we also write $h : \left(A,\alpha\right) \rightarrow \left(B, \beta\right)$.

A special $F$-algebra is the \emph{free $F$-algebra} 
\begin{align*}
  F^{\star} = (\mu F, \cons_F)
\end{align*}
for which there is a homomorphism
\begin{align*}
  \kata{\alpha} : F^{\star} \rightarrow \left(A, \alpha\right)
\end{align*}
for any other algebra $\left(A, \alpha\right)$.
We call those homomorphisms $\kata{\alpha}$ \emph{$F$-algebra catamorphisms}.
The functions $\kata{\alpha}$ encapsulate structured recursion on values in $\mu F$ with the semantics defined by the function $\alpha$, which is only defined on flat terms.
The carrier of $F^{\star}$, with $\mu F = F \mu F$, is the set of inductively defined values in the shape defined by $F$.
The function $\cons_F : F \mu F \rightarrow \mu F$ inductively constructs the values in $\mu F$.

$F$-algebras and especially $F$-catamorphisms give us a tool to map the abstractly defined syntax of FreeCHR (the free $F$-algebra) to concrete implementations (other $F$-algebras).
By this, we have a strong link between theoretical definitions and actual implementations, which allows us to define theorems on and prove them along the inductive structure of the formal definition.

\subsection{Labelled transition systems}
A \emph{labelled transition system} (LTS) $\omega = \langle \Sigma, L, (\mapsto) \rangle$
consists of a set $\Sigma$ called the \emph{domain},
a set $L$ called the \emph{labels}
and a ternary \emph{transition relation} $R \subseteq \Sigma \times L \times \Sigma$.
The idea is that if $s \xmapsto{\ l\ } s' \in R$, we transition from state $s$ to $s'$ by the action $l$.

For two LTS $\omega_1 = \langle \Sigma_1, L_1, (\mapsto) \rangle$ and $\omega_2 = \langle \Sigma_2, L_2, (\hookrightarrow) \rangle$
and a functions $f : \Sigma_1 \longrightarrow \Sigma_2$ we say that $\omega_1$ is $(f, g)$-sound \WRT $\omega_2$, \IFF
\begin{align*}
  s \xmapsto{\ l\ } s' \in (\mapsto) \Longrightarrow f(s) \xhookrightarrow{\ l\ } f(s') \in (\hookrightarrow)\tag{$f$-soundness}
\end{align*}
By $(\mapsto)^{+}$ we denote the \emph{transitive} and by $(\mapsto)^{*}$ the \emph{reflexive-transitive} closure of $(\mapsto)$.
Recall that $(\mapsto)^{+} \subset (\mapsto)^{*}$, for every $R \subseteq (\mapsto)^{+}$, $R^{+} \subseteq (\mapsto)^{+}$ and
for every $Q \subseteq (\mapsto)^{*}$, $Q^{*} \subseteq (\mapsto)^{*}$.
  \section{Constraint Handling Rules}\label{sec:non-herbrand-chr}
In this section, we want to give an informal introduction to \emph{Constraint Handling Rules} (CHR) and its \emph{refined} operational semantics $\omega_r$.

\subsection{Syntax}
CHR is an embedded rule-based programming language rules of the generalized form
\begin{align*}
  N\ @\ K\ \setminus\ R\ \Longleftrightarrow\ \left[G\ |\right]\ B
\end{align*}
$N$ is the \emph{unique name} of the rule.
$K$ is called the \emph{kept} and $R$ the \emph{removed head}.
Both are sequences of patterns over the domain of values.
Either can be omitted, but not both at the same time.
If $K$ is empty, we call the rule a \emph{simplification} rule.
If $R$ is empty, we call it a \emph{propagation} rule and write them with $(\Longrightarrow)$ instead of $(\Longleftrightarrow)$.
$G$ is an optional condition called the \emph{guard}.
If $G$ is omitted, we assume that it is true.
$B$ the \emph{body} of the rule, which is a sequence of values.

\begin{example}[Greatest common divisor]\label{ex:chr:gcd}
  The program in 
  \begin{align*}
    \mathit{zero}\ &@\ 0\ \Longleftrightarrow\ \eps\\
    \mathit{subtract}\ &@\ N\ \setminus\ M\ \Longleftrightarrow\ 0 < N \band 0 < M \band N \leq M\ |\ M-N
  \end{align*}
  implements the Euclidean algorithm to compute the greatest common divisor of a collection of numbers.
  The rule \emph{zero} removes any numbers equal to $0$.
  The rule \emph{subtract} replaces $M$ by the difference $M-N$ for pairs of numbers $N$ and $M$ with $0 < N, M$ and $N \leq M$.
\end{example}
We also need the concept of an \emph{instance} of a rule.
We assume that any expressions (like $N-M$) are evaluated according to the semantics provided by the host language.
We will write $e \equiv c$, if the expression $e$ evaluates to $c$.
In abstract examples, we use the intuitive meaning of operators and functions.
\begin{definition}[Ground evaluated rule instances]
  Given a rule 
  \begin{align*}
    r = \left(N\ @\ K\ \setminus\ R\ \Longleftrightarrow\ \left[G\ |\right]\ B\right)
  \end{align*}
  and a substitution $\theta$, we call $r\theta$ a \emph{ground instance} of $r$, \IFF $\theta$ substitutes all variables in $r$ by a ground value and all host-language expressions in $r\theta$ are evaluated according to the semantics, provided by the host language.
\end{definition}

\begin{example}[Ground evaluated instance of \textit{subtract}]
  With $\theta = \left\{N \mapsto 9, M \mapsto 12\right\}$
  \begin{align*}
    \mathit{subtract}\ &@\ 9\ \setminus\ 12\ \Longleftrightarrow\ \btrue\ |\ 3
  \end{align*}
  is a \emph{ground evaluated instance} of the \textit{subtract} rule of \autoref{ex:chr:gcd}.
\end{example}

\subsection{Refined operational semantics}
The very abstract operational semantics of CHR operates on plain multisets of values and describes that a rule can fire if for each pattern of the rule there is a unique matching value in the multiset and if the guard holds for the found values.
How those values are found and which rules are tried to be applied is nondeterministic \cite{fruehwirth2009constraint}.
The refined operational semantics formalize how the program is traversed in order to find a rule and how the head of the rule is traversed in order to find a suitable matching.
We want to give a semi-formal introduction for the refined operational semantics for \emph{ground} CHR.

\subsubsection{States}
The refined operational semantics operates on states of the form $\langle Q, S, H, I \rangle$.

\paragraph{Query}
The \emph{sequence} $Q$ is called the \emph{query} and simulates a call stack.
Values on the query may be interpreted as pending or running procedure calls.
The query is the main driver for execution of CHR programs.
The value on top is the currently \emph{active} value.
The values on the query are decorated in several ways, which we will discuss later.

\paragraph{Store}
The \emph{set} $S$ is called the \emph{store} and contains activated values, decorated with a unique index.
The index is used to simulate a multiset, as required by the very abstract operational semantics, as well as for other purposes which we will also discuss later.

\paragraph{Propagation history}
The \emph{set} $H$ is called the \emph{propagation history}.
It contains tuples of rule names and indices which refer to values in the store.
We will call those tuples \emph{configurations}, \emph{history entries} or \emph{records} later on.
Those entries record which rules were applied to which values.
By checking the history upon rule application, repeated application of the same rule on the same values, and hence non-termination, is prevented.

\paragraph{Index}
The \emph{integer} $I$ is called the \emph{index}.
Every time a value is activated, the current value of $I$ is used as its identifier and $I$ is incremented.
Hereby we generate unique identifiers for newly activated values.

\subsubsection{Programs}
Programs for the refined operational semantics are \emph{sequences} of rules.
The head patterns of the rules of a program are viewed as decorated with indices descending from \emph{right to left} and \emph{top to bottom} (in textual order) throughout the program.
We call them \emph{pattern indices}.

\begin{example}[Greatest common divisor decorated]\label{ex:chr:gcd:decorated}
  The program 
  \begin{align*}
    \mathit{zero}\ &@\ 0^{\#1}\ \Longleftrightarrow\ \emptyset\\
    \mathit{subtract}\ &@\ N^{\#3}\ \setminus\ M^{\#2}\ \Longleftrightarrow\ 0 < N \band 0 < M \band N \leq M\ |\ M-N
  \end{align*}
  shows the program in \autoref{ex:chr:gcd} with indexed head patterns.
\end{example}

\subsubsection{Transitions}
The original definition as stated by Frühwirth \cite{fruehwirth2009constraint} describes six kinds of state transitions.
Since we operate on ground values we can ignore two of them which are only concerned with non-ground values.

\paragraph{Activate}
The transition
\begin{align*}
  \langle c:Q, S, H, I \rangle
  \xmapsto{\textsc{activate}} \langle (I, c)^{\#1} : Q, \left\{\left(I, c\right)\right\} \cup S, H, I+1 \rangle
\end{align*}
"calls" a value $c$ by introducing it to the store with a fresh index $I$.
On the query, the value is also decorated with the pattern index $\#1$.
This indicates that it will be tried to match it to the rightmost pattern of the first rule.
We use Haskell-like syntax to denote lists:
$[]$ is the empty list and $\mathit{x}:\mathit{xs}$ constructs a list with head element $\mathit{x}$ and tail $\mathit{xs}$.
We will use the notations $\left(a : b : c : []\right)$ and $\left[a, b, c\right]$ interchangeably as we consider it useful.

\paragraph{Apply}
Given a ground evaluated instance
\begin{align*}
  r\ &@\ c_1^{\#l_1}, ..., c_{k}^{\#l_k}\ \setminus\ c_{k+1}^{\#l_k+1}, ..., c_{n}^{l_n}\ \Longleftrightarrow\ \btrue\ |\ B'
\end{align*}
of a rule
\begin{align*}
  r\ &@\ h_1^{\#l_1}, ..., h_{k}^{\#l_k}\ \setminus\ h_{k+1}^{\#l_k+1}, ..., h_{n}^{l_n}\ \Longleftrightarrow\ G\ |\ B
\end{align*}
such that for $1 \leq j \leq n$, \nolinebreak{$K=\left\{\left(i_1, c_1\right), ..., \left(i_k, c_k\right)\right\}$} and \linebreak\nolinebreak{$R=\left\{\left(i_{k+1}, c_{k+1}\right), ..., \left(i_{n}, c_{n}\right)\right\}$}, $(i_j, c_j) \in K \cup R$ and \nolinebreak{$\left\{(r, i_1, ..., i_n)\right\} \notin H$}, we can perform the transition
\begin{align*}
  &\langle (i_j, c_j)^{\#l_j} : Q, K \uplus R \uplus S, H, I\rangle\\
  &\quad\xmapsto{\textsc{apply}} \langle B' \diamond ((i_j, c_j)^{\#l_j} : Q), K \uplus S, \left\{(r, i_1, ..., i_n)\right\} \cup H, I \rangle
\end{align*}
We use the operator $(\uplus)$ on sets to emphasize, that the operand sets are disjoint, \IE $A \uplus B = C$ \IFF $A \cup B = C$ and $A \cap B = \emptyset$.

Note, that the indices $l_p$ increment from right to left, \IE $l_p = l_{p+1} + 1$.
We can apply the transition if $c_j$ matches the ${l_j}^{\text{th}}$ pattern of the head and for each other pattern $h_{\iota}$, there is a $(i_{\iota}, c_{\iota})$ in the store.
We need to check if the configuration $(r, i_1, ..., i_n)$ already fired, to prevent possible repeated application.
If not, we record the configuration, remove $R$ from the store and query the values of the body, by concatenating the sequence $B'$ before the query.
The operator $(\diamond)$ denotes concatenation of two sequences, \IE $\left[a_1, ..., a_n\right] \diamond \left[b_1, ..., b_m\right] = \left[a_1, ..., a_n, b_1, ..., b_m\right]$.

\paragraph{Drop}
The transition
\begin{align*}
  \langle (i, c)^{\#j} : Q, S, H, I \rangle \xmapsto{\textsc{drop}} \langle Q, S, H, I \rangle
\end{align*}
is used if $j$ exeeds the pattern indices of the program. This indicates that there are no more applicable rules for the currently active value.
This also happens if $(i,c)$ was removed from the store by the \textsc{apply} transition at some point.

\paragraph{Default}
If no other transition is possible
\begin{align*}
  \langle (i, c)^{\#j} : Q, S, H, I \rangle \xmapsto{\textsc{default}} \langle (i, c)^{\#j+1} : Q, S, H, I \rangle 
\end{align*}
is used. This transition continues the traversal with the currently active value through the program by incrementing the pattern index.
Thereby, it will be attempted to match $(i,c)$ to the next pattern in the program. 

\subsubsection{Execution}
Given a program and an initial state $\langle Q, \emptyset, \emptyset, 1\rangle$, the transition rules described above are applied until $Q$ is empty.

We want to show on two examples, how execution of the refined operational semantics works.

\begin{figure}
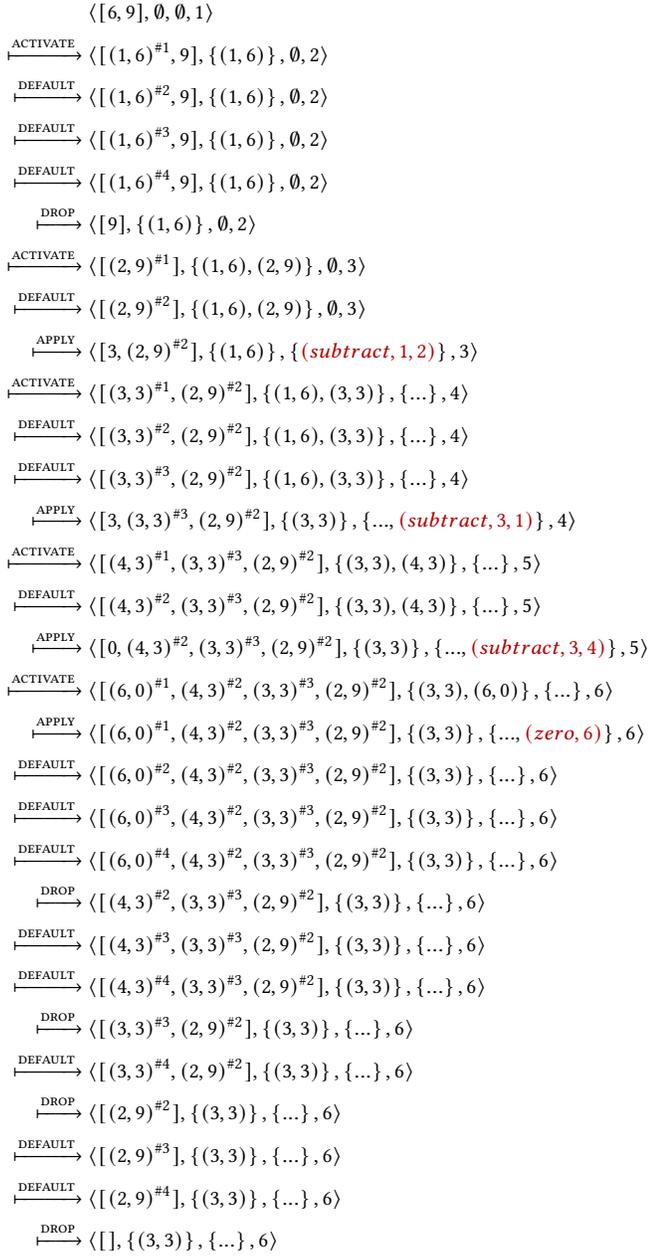

  {\small\begin{align*}
                        &\ \langle [6, 9], \emptyset, \emptyset, 1 \rangle \\
    \xmapsto{\textsc{activate}}  &\ \langle [(1, 6)^{\#1}, 9], \left\{(1, 6)\right\}, \emptyset, 2 \rangle\\
    \xmapsto{\textsc{default}}   &\ \langle [(1, 6)^{\#2}, 9], \left\{(1, 6)\right\}, \emptyset, 2 \rangle \\
    \xmapsto{\textsc{default}}   &\ \langle [(1, 6)^{\#3}, 9], \left\{(1, 6)\right\}, \emptyset, 2 \rangle \\
    \xmapsto{\textsc{default}}   &\ \langle [(1, 6)^{\#4}, 9], \left\{(1, 6)\right\}, \emptyset, 2 \rangle \\
    \xmapsto{\textsc{drop}}      &\ \langle [9], \left\{(1, 6)\right\}, \emptyset, 2 \rangle \\
    \xmapsto{\textsc{activate}}  &\ \langle [(2, 9)^{\#1}], \left\{(1, 6), (2, 9)\right\}, \emptyset, 3 \rangle \\
    \xmapsto{\textsc{default}}   &\ \langle [(2, 9)^{\#2}], \left\{(1, 6), (2, 9)\right\}, \emptyset, 3 \rangle \\
    \xmapsto{\textsc{apply}}     &\ \langle [3, (2, 9)^{\#2}], \left\{(1, 6)\right\}, \left\{\textcolor{red!70!black}{(subtract, 1, 2)}\right\}, 3 \rangle \\
    \xmapsto{\textsc{activate}}  &\ \langle [(3, 3)^{\#1}, (2, 9)^{\#2}], \left\{(1, 6), (3, 3)\right\}, \left\{...\right\}, 4 \rangle \\
    \xmapsto{\textsc{default}}   &\ \langle [(3, 3)^{\#2}, (2, 9)^{\#2}], \left\{(1, 6), (3, 3)\right\}, \left\{...\right\}, 4 \rangle \\
    \xmapsto{\textsc{default}}   &\ \langle [(3, 3)^{\#3}, (2, 9)^{\#2}], \left\{(1, 6), (3, 3)\right\}, \left\{...\right\}, 4 \rangle \\
    \xmapsto{\textsc{apply}}     &\ \langle [3, (3, 3)^{\#3}, (2, 9)^{\#2}], \left\{(3, 3)\right\}, \left\{..., \textcolor{red!70!black}{(subtract, 3, 1)}\right\}, 4 \rangle \\
    \xmapsto{\textsc{activate}}  &\ \langle [(4, 3)^{\#1}, (3, 3)^{\#3}, (2, 9)^{\#2}], \left\{(3, 3), (4, 3)\right\}, \left\{...\right\}, 5 \rangle \\
    \xmapsto{\textsc{default}}   &\ \langle [(4, 3)^{\#2}, (3, 3)^{\#3}, (2, 9)^{\#2}], \left\{(3, 3), (4, 3)\right\}, \left\{...\right\}, 5 \rangle \\
    \xmapsto{\textsc{apply}}     &\ \langle [0, (4, 3)^{\#2}, (3, 3)^{\#3}, (2, 9)^{\#2}], \left\{(3, 3)\right\}, \left\{...,\textcolor{red!70!black}{(subtract, 3, 4)}\right\}, 5 \rangle \\
    \xmapsto{\textsc{activate}}  &\ \langle [(6, 0)^{\#1}, (4, 3)^{\#2}, (3, 3)^{\#3}, (2, 9)^{\#2}], \left\{(3, 3), (6, 0)\right\}, \left\{...\right\}, 6 \rangle \\
    \xmapsto{\textsc{apply}}     &\ \langle [(6, 0)^{\#1}, (4, 3)^{\#2}, (3, 3)^{\#3}, (2, 9)^{\#2}], \left\{(3, 3)\right\}, \left\{...,\textcolor{red!70!black}{(zero, 6)}\right\}, 6 \rangle \\
    \xmapsto{\textsc{default}}   &\ \langle [(6, 0)^{\#2}, (4, 3)^{\#2}, (3, 3)^{\#3}, (2, 9)^{\#2}], \left\{(3, 3)\right\}, \left\{...\right\}, 6 \rangle \\
    \xmapsto{\textsc{default}}   &\ \langle [(6, 0)^{\#3}, (4, 3)^{\#2}, (3, 3)^{\#3}, (2, 9)^{\#2}], \left\{(3, 3)\right\}, \left\{...\right\}, 6 \rangle \\
    \xmapsto{\textsc{default}}   &\ \langle [(6, 0)^{\#4}, (4, 3)^{\#2}, (3, 3)^{\#3}, (2, 9)^{\#2}], \left\{(3, 3)\right\}, \left\{...\right\}, 6 \rangle \\
    \xmapsto{\textsc{drop}}      &\ \langle [(4, 3)^{\#2}, (3, 3)^{\#3}, (2, 9)^{\#2}], \left\{(3, 3)\right\}, \left\{...\right\}, 6 \rangle \\
    \xmapsto{\textsc{default}}   &\ \langle [(4, 3)^{\#3}, (3, 3)^{\#3}, (2, 9)^{\#2}], \left\{(3, 3)\right\}, \left\{...\right\}, 6 \rangle \\
    \xmapsto{\textsc{default}}   &\ \langle [(4, 3)^{\#4}, (3, 3)^{\#3}, (2, 9)^{\#2}], \left\{(3, 3)\right\}, \left\{...\right\}, 6 \rangle \\
    \xmapsto{\textsc{drop}}      &\ \langle [(3, 3)^{\#3}, (2, 9)^{\#2}], \left\{(3, 3)\right\}, \left\{...\right\}, 6 \rangle \\
    \xmapsto{\textsc{default}}   &\ \langle [(3, 3)^{\#4}, (2, 9)^{\#2}], \left\{(3, 3)\right\}, \left\{...\right\}, 6 \rangle \\
    \xmapsto{\textsc{drop}}      &\ \langle [(2, 9)^{\#2}], \left\{(3, 3)\right\}, \left\{...\right\}, 6 \rangle \\
    \xmapsto{\textsc{default}}   &\ \langle [(2, 9)^{\#3}], \left\{(3, 3)\right\}, \left\{...\right\}, 6 \rangle \\
    \xmapsto{\textsc{default}}   &\ \langle [(2, 9)^{\#4}], \left\{(3, 3)\right\}, \left\{...\right\}, 6 \rangle \\
    \xmapsto{\textsc{drop}}      &\ \langle [], \left\{(3, 3)\right\}, \left\{...\right\}, 6 \rangle
  \end{align*}}
  \caption{Execution of the Euclidean algorithm implemented in CHR with initial query $[6,9]$}
  \label{fig:chr:gcd:exec}
\end{figure}

\begin{example}[Greatest common divisor executed]\label{ex:chr:gcd:exec}
  \autoref{fig:chr:gcd:exec} demonstrates the refined operational semantics on the example query $\left[6, 9\right]$ and the Euclidean algorithm program of \autoref{ex:chr:gcd:decorated}.
  Since the program does not contain any propagation rules, we will abbreviate the propagation history and only show it to emphasize which rule was applied to which values.

  First, the value $6$ is activated and introduced to the store.
  Since there are no other values in the store yet, its pattern index gets incremented until it is dropped.
  Then, the value $9$ is activated and its pattern index gets incremented once.
  It now matches the pattern $M^{\#2}$ of the rule $subtract$ and with $6$ matched on $N^{\#3}$, the guard evaluates to $\btrue$ as well.
  Hence, the value $9-6 = 3$ is queried and $(2, 9)$ removed from the store.
  Effectively replacing $9$ with $9-6$.
  The value gets activated and its pattern index incremented to $3$, where the rule $subtract$ can be applied again.
  This time, $6-3 = 3$ is queried and $(1, 6)$ removed, replacing $6$ with $6-3$.
  Now again, $3$ is activated with index $4$ and after one \textsc{default} transition the rule $subtract$ fires again, replacing this newly added $3$ with $0$.
  $0$ then gets activated and instantly matches the $0^{\#1}$ pattern.
  Hence, the rule \emph{zero} fires and removes the value $(6,0)$ from the store.
  At this point, all values except $(3,3)$ are no longer alive and no more non-active values are on the query.
  Hence, all values are successively dropped from the query and the execution finally terminates.
\end{example}

\begin{figure*}[t]
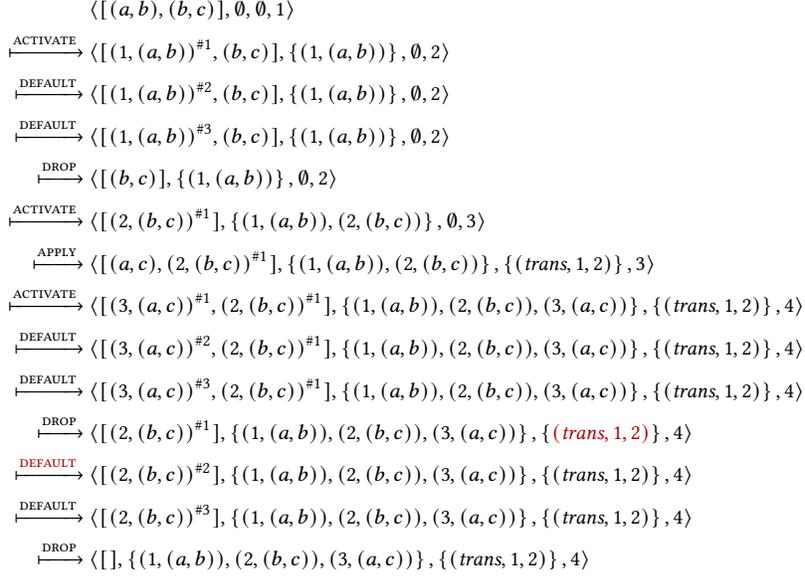

  \begin{center}
    \begin{minipage}{.7\textwidth}
      {\small
      \begin{align*}
                                            &\ \langle [(a, b), (b, c)], \emptyset, \emptyset, 1 \rangle \\
        \xmapsto{\textsc{activate}}         &\ \langle [(1, (a, b))^{\#1}, (b, c)], \left\{(1, (a, b))\right\}, \emptyset, 2 \rangle \\
        \xmapsto{\textsc{default}}          &\ \langle [(1, (a, b))^{\#2}, (b, c)], \left\{(1, (a, b))\right\}, \emptyset, 2 \rangle \\
        \xmapsto{\textsc{default}}          &\ \langle [(1, (a, b))^{\#3}, (b, c)], \left\{(1, (a, b))\right\}, \emptyset, 2 \rangle \\
        \xmapsto{\textsc{drop}}             &\ \langle [(b, c)], \left\{(1, (a, b))\right\}, \emptyset, 2 \rangle \\
        \xmapsto{\textsc{activate}}         &\ \langle [(2, (b, c))^{\#1}], \left\{(1, (a, b)), (2, (b, c))\right\}, \emptyset, 3 \rangle \\
        \xmapsto{\textsc{apply}}            &\ \langle [(a, c), (2, (b, c))^{\#1}], \left\{(1, (a, b)), (2, (b, c))\right\}, \left\{(\mathit{trans}, 1, 2)\right\}, 3 \rangle \\
        \xmapsto{\textsc{activate}}         &\ \langle [(3, (a, c))^{\#1}, (2, (b, c))^{\#1}], \left\{(1, (a, b)), (2, (b, c)), (3, (a, c))\right\}, \left\{(\mathit{trans}, 1, 2)\right\}, 4 \rangle \\
        \xmapsto{\textsc{default}}          &\ \langle [(3, (a, c))^{\#2}, (2, (b, c))^{\#1}], \left\{(1, (a, b)), (2, (b, c)), (3, (a, c))\right\}, \left\{(\mathit{trans}, 1, 2)\right\}, 4 \rangle \\
        \xmapsto{\textsc{default}}          &\ \langle [(3, (a, c))^{\#3}, (2, (b, c))^{\#1}], \left\{(1, (a, b)), (2, (b, c)), (3, (a, c))\right\}, \left\{(\mathit{trans}, 1, 2)\right\}, 4 \rangle \\
        \xmapsto{\textsc{drop}}             &\ \langle [(2, (b, c))^{\#1}], \left\{(1, (a, b)), (2, (b, c)), (3, (a, c))\right\}, \left\{\textcolor{red!70!black}{(\mathit{trans}, 1, 2)}\right\}, 4 \rangle \\
        \xmapsto{\textcolor{red!70!black}{\textsc{default}}} &\ \langle [(2, (b, c))^{\#2}], \left\{(1, (a, b)), (2, (b, c)), (3, (a, c))\right\}, \left\{(\mathit{trans}, 1, 2)\right\}, 4 \rangle \\
        \xmapsto{\textsc{default}}          &\ \langle [(2, (b, c))^{\#3}], \left\{(1, (a, b)), (2, (b, c)), (3, (a, c))\right\}, \left\{(\mathit{trans}, 1, 2)\right\}, 4 \rangle \\
        \xmapsto{\textsc{drop}}             &\ \langle [], \left\{(1, (a, b)), (2, (b, c)), (3, (a, c))\right\}, \left\{(\mathit{trans}, 1, 2)\right\}, 4 \rangle \\
      \end{align*}}
    \end{minipage}
  \end{center}
  \caption{Demonstration of the effect of the propagation history.}
  \label{fig:chr:trans}
\end{figure*}

\begin{example}[Transitive hull]\label{ex:chr:trans}
  The program
  \begin{align*}
    \mathit{trans}\ @\ \left(X, Y\right)^{\#2},\ \left(Y, Z\right)^{\#1}\ \Longrightarrow\ X \neq Z\ |\ \left(X, Z\right)
  \end{align*}
  adds the transitive edge $(X, Z)$ of two edges $(X,Y)$ and $(Y, Z)$, with $X \neq Z$.
  \autoref{fig:chr:trans} shows the execution of the program with an initial query $[(a, b), (b, c)]$.

  First, $(a, b)$ gets activated and after a two \textsc{default} transitions dropped, as the rule requires two values to fire.
  Then, $(b, c)$ gets activated and the rule \emph{trans} fires immediately.
  This queries $(a, c)$ and adds the record $(\textit{trans}, 1, 2)$ to the propagation history.
  Since there is no matching partner for $(a, c)$ in the store, the value gets dropped after activation and two \textsc{default} transitions.
  
  Now, $(b, c)$ is active again.
  Without the propagation history, the rule from above could be applied again, as both $(1, (a, b))$ and $(2, (b, c))$ are still alive.
  But since the record $(\textit{trans}, 1, 2)$ is already within the propagation history the \textsc{apply} transition can not be applied.
  Hence, the \textsc{default} transition needs to be applied and the value is dropped ultimately.
\end{example}
  \section{FreeCHR}\label{sec:freechr}
FreeCHR was introduced as a framework to embed CHR into arbitrary programming languages.
The main idea is to model the syntax of programs as an endofunctor within the domain of the host language.
We want to briefly reiterate the necessary definitions and refer the reader to the original publication \cite{rechenberger2023freechr} for further details.

\begin{definition}[Syntax of FreeCHR programs]\label{def::freechr:syn:set:functor}
    The functor
    \begin{align*}
        \chr_C D &=\tstr \times \flist  \tbool^C \times \flist  \tbool^C \times \tbool^{\flist  C} \times (\flist C)^{\flist  C} \\
            &\sqcup D \times D
    \end{align*}
    describes the syntax of FreeCHR programs.
\end{definition}
  
The set $\tstr \times \flist \tbool^C \times \flist \tbool^C \times \tbool^{\flist  C} \times (\flist C)^{\flist C}$ is the set of single rules.
The name of the rule is a string in $\tstr$.
The kept and removed head are sequences of functions in $\flist \tbool^C$ which map elements of $C$ to Booleans, effectively checking individual values for applicability of the rule.
The guard is a function in $\tbool^{\flist C}$ and maps sequences of elements in $C$ to Booleans, checking all matched values in the context.
Finally, the body is a function in $(\flist C)^{\flist C}$ and maps the matched values to a sequence of newly generated values.

The set $D \times D$ represents the composition of FreeCHR programs by an execution strategy, allowing the construction of more complex programs from, ultimately, single rules.

By the structure of $\chr_C$, a $\chr_C$-algebra with carrier $D$ is defined by two functions 
\begin{align*}
\rho &:\ \tstr \times \flist \tbool^C \times \flist \tbool^C \times \tbool^{\flist C} \times (\flist C)^{\flist C} \longrightarrow D\\
\nu &:\ D \times D \rightarrow D 
\end{align*}
as $(D, \left[\rho, \nu\right])$.
A $\chr_C$-algebra is called an \emph{instance} of FreeCHR.
The free $\chr_C$-algebra
\begin{align*}
    \chrstar = \left(\mu\chr_C, \left[\chrrule, \chrcomp\right]\right)
\end{align*}
with
\begin{align*}
    \mu\chr_C
        &=\tstr \times \flist \tbool^C \times \flist \tbool^C \times \tbool^{\flist C} \times (\flist C)^{\flist C}\\
        &\sqcup \mu\chr_C \times \mu\chr_C
\end{align*}
and injections
\begin{align*}
    rule &: \tstr \times \flist \tbool^C \times \flist \tbool^C \times \tbool^{\flist C} \times (\flist C)^{\flist C} \\
         &\quad\longrightarrow \mu\chr_C \\
    \chrcomp &: \mu\chr_C \times \mu\chr_C \longrightarrow \mu\chr_C
\end{align*}
provides us with an inductively defined representation of programs.
We use the $\chr_C$-catamorphisms
\begin{align*}
    \kata{\left[\rho,\nu\right]} &: \mu\chr_C \longrightarrow D\\
    \kata{\left[\rho, \nu\right]}(rule(n,k,r,g,b)) &= \rho(n,k,r,g,b)\\
    \kata{\left[\rho, \nu\right]}(p_1 \chrcomp ... \chrcomp p_n) &= \nu(\kata{\left[\rho, \nu\right]}(p_1), ...,\kata{\left[\rho, \nu\right]}(p_n))
\end{align*}
to map inductively defined programs in $\chrstar$ to programs in an instance $(D, \left[\rho, \nu\right])$ of FreeCHR.
  
The program from \autoref{ex:chr:gcd} can be expressed in FreeCHR as shown in \autoref{ex:freechr:gcd}.

\begin{example}[Euclidean algorithm (cont.)]\label{ex:freechr:gcd}
    The program 
    \begin{align*}
        \mathit{gcd} &= \mathit{zero} \chrcomp \mathit{subtract}
    \end{align*}
    with
    \begin{align*}
        \mathit{zero} &= rule(\mathtt{"zero"}, \eps, \left[\lambda n. n = 0\right], (\lambda n. \btrue), (\lambda n. \eps)) \\
        \mathit{subtract} &= rule(\mathtt{"subtract"}, \left[\lambda n. 0 < n\right], \left[\lambda m. 0 < m\right],\\
            &\quad\quad (\lambda n\ m. n \leq m), (\lambda n\ m. \left[m-n\right]))
    \end{align*}
    implements the Euclidean algorithm, as defined in \autoref{ex:chr:gcd}.
    $\lambda$-abstractions are used for ad-hoc definitions of functions.
    To reduce formal clutter, we write the functions of guard and body as $n$-ary functions instead of unary functions, \IE $(\lambda n\ m. n \leq m)$ instead of $(\lambda \left[n,m\right]. n \leq m)$.
\end{example}

Finally, we want to recall the \emph{very abstract} operational semantics $\omstara$ of FreeCHR originally defined in \cite{rechenberger2023freechr}.

\begin{definition}[\emph{Very abstract} operational semantics $\omstara$]\label{def::freechr:sem:op:very-abstract}
    The \emph{very abstract} operational semantics of FreeCHR is defined as the labelled transition system
    \begin{align*}
        \omstara = \langle \fmultiset C, \mu\chr_C, (\freetr{})^{*} \rangle
    \end{align*} where the transition relation $(\freetr{}) \subset \fmultiset C \times \mu\chr_C \times \fmultiset C$ is defined by the inference rules below.
    The functor $\fmultiset$ maps a set $X$ to the set $\fmultiset X$ of multisets over $X$.

    \subparagraph*{\textbf{Rule selection}}
    The transition
    \begin{align*}
        \begin{prooftree}[small]
            \hypo{s \freetr{p_j} s'}
            \infer1[step]{s \freetr{p_1 \chrcomp ... \chrcomp p_j \chrcomp ... \chrcomp p_l} s'}
        \end{prooftree}
    \end{align*}
    selects a component program $p_j$ from the composite program $p_1 \chrcomp ... \chrcomp p_j \chrcomp ... \chrcomp p_l$.

    \subparagraph*{\textbf{Rule application}}
    The transition
    \begin{align*}
        \begin{prooftree}[small]
            \hypo{k_1(c_1) \band ... \band k_{n}(c_{n}) \band r_{1}(c_{n+1}) \band ... \band r_{m}(c_{n+m}) \band g(c_1,...,c_{n+m}) \equiv_{\tbool} \btrue}
            \infer1[apply]{\left\{c_1,  ...,  c_{n+m}\right\} \uplus \Delta S \freetr{\chrrule(N, k, r, g, b)} \left\{c_1, ..., c_{n}\right\} \uplus b(c_1,...,c_{n+m}) \uplus \Delta S}
        \end{prooftree}
    \end{align*}
    where $k = \left[k_1, ..., k_{n}\right]$ and $r = \left[r_{1}, ..., r_{m}\right]$, 
    applies a rule to the current state of the program if the state contains a unique value for each pattern in the head of the rule and these values satisfy the guard.
\end{definition}

The \textsc{step} transition models the selection of a \emph{subprogram} $p_j$ from the composite program $p_1 \chrcomp ... \chrcomp p_l$.
If we are able to perform a transition with $p_j$, we are able to perform it with $p_1 \chrcomp ... \chrcomp p_l$ as well.
\textsc{apply} states that a rule can be applied if there is a value for every pattern in the head that satisfy the guard of the program.
If the rule is applied, the values ${c_{n+1}, ...,c_{n+m}}$ (which were matched to the removed head) are removed from the store and the values generated by the body are added.

  \section{States for the refined operational semantics}\label{sec:ExtendedStates}
We now want to model the structure of the states required for implementing the \emph{refined} operational semantics.

\begin{definition}[States]\label{def:states}
    The \Set-endofunctor
    \begin{align*}
        \Omega_r C =&\ \flist ((\tnat \setsum \tone) \setprod C) \tag{Query}\\
            \setprod&\ \powerset (\tnat \setprod C) \tag{Store}\\
            \setprod&\ \powerset (\tstr \setprod \flist \tnat) \tag{Propagation history}\\
            \setprod&\ \tnat \tag{Index}
    \end{align*}
    models the set of states over values $C$.
    For an Element $\langle Q, S, H, I \rangle \in \Omega_r C$ we call
    $Q$ the \emph{query}, $S$ the \emph{store}, $H$ the \emph{propagation history} and $I$ the \emph{index}.

    Furthermore, we define the projections defined in \autoref{fig:state:projections} to extract the elements of a state.
    \begin{figure}[t]
        \begin{align*}
            \mathtt{query}_C&: \Omega_r C \longrightarrow \flist (\left(\tnat \setsum \tone\right) \setprod C)\\
            \mathtt{query}_C&\langle Q, \_, \_, \_ \rangle = Q
        \end{align*}

        \begin{align*}
            \mathtt{store}_C&: \Omega_r C \longrightarrow \powerset (\tnat \setprod C)\\
            \mathtt{store}_C&\langle \_, S, \_, \_ \rangle = S
        \end{align*}

        \begin{align*}
            \mathtt{history}_C&: \Omega_r C \longrightarrow \powerset (\tstr \setprod \flist \tnat)\\
            \mathtt{history}_C&\langle \_, \_, H, \_ \rangle = H
        \end{align*}

        \begin{align*}
            \mathtt{index}_C&: \Omega_r C \longrightarrow \tnat \\
            \mathtt{index}_C&\langle \_, \_, \_, I \rangle = I
        \end{align*}
        \caption{Projections on elements of $\Omega_r C$.}
        \label{fig:state:projections}
    \end{figure}
\end{definition}
The states are modelled as discussed in \autoref{sec:non-herbrand-chr} and the components serve the same functions.
Additional to the projection functions, we will define functions to modify the state.

\begin{definition}[Operations on the \emph{query}]\label{def:states:query}
    The functions in \autoref{fig:states:query} define operations on the \emph{query} of a state.
    \begin{figure}[t]
        \begin{align*}
            \mathtt{push\_query}_C&: \Omega_r C \setprod \flist C \longrightarrow \Omega_r C \\
            \mathtt{push\_query}_C&\left(\langle Q, S, H, I\rangle, c_1, ..., c_n\right) =\\
                &\quad \langle \left(\bot, c_1\right):...:\left(\bot, c_n\right):Q, S, H, I \rangle
        \end{align*}

        \begin{align*}
            \mathtt{pop\_query}_C& : \Omega_r C \longrightarrow \Omega_r C\\
            \mathtt{pop\_query}_C&\langle (i, c):Q, S, H, I\rangle = \langle Q, S, H, I\rangle
        \end{align*}
        \caption{Operations on the \emph{query} of a state}
        \label{fig:states:query}
    \end{figure}
\end{definition}

The functions in \autoref{fig:states:query} define basic stack operations on the query of a state.
Note, that $\mathtt{pop\_query}_C$ is a partial function, defined only if the query is not empty.
It is up to the implementor, to handle the undefined case (\IE throw an exception or stay with the default handling provided by the programming language).
The symbol $\bot$ signals that a value was not yet activated.

\begin{definition}[Operations on the \emph{store}]\label{def:states:store}
    The functions in \autoref{fig:states:store} define operations on the \emph{store} of a state.
    \begin{figure}[t]
        \begin{align*}
            \mathtt{activate}_C& : \Omega_r C \longrightarrow \tnat \times \Omega_r C\\
            \mathtt{activate}_C&\left(\langle (\bot, c):Q, S, H, I\rangle, c\right) =\\
                &\quad\left(I, \langle (I, c):Q, S \cup \left\{(I, c)\right\}, H, I+1\rangle\right)
        \end{align*}

        \begin{align*}
            \mathtt{remove}_C& : \Omega_r C \setprod \flist\tnat \longrightarrow \Omega_r C\\
            \mathtt{remove}_C&\left(\langle Q, S \cup \left\{(i_1, c_1), ..., (i_n, c_n)\right\}, H, I\rangle, i_1, ..., i_n\right) =\\
                &\quad\langle Q, S, H, I\rangle
        \end{align*}

        \begin{align*}
            \mathtt{alive}_C& : \Omega_r C \setprod \tnat \longrightarrow \tbool\\
            \mathtt{alive}_C&\left(\langle Q, S, H, I\rangle, i\right) = \left\{\begin{matrix}
                \btrue &  & \exists c \in C. (i, c) \in S\\
                \bfalse &  & \text{otherwise}
            \end{matrix}
            \right.
        \end{align*}
        \caption{Operations on the \emph{store} of a state}
        \label{fig:states:store}
    \end{figure}
\end{definition}

The function $\mathtt{activate}_C$ adds the top value of the \emph{query} to the \emph{store}, with the current \emph{index} as the unique identifier and increments it by $1$ if this value was not activated before.
It also replaces the value with a decorated version.
If the value is already active or the query empty, the function is not defined.
$\mathtt{remove}_C$ removes the values with the given identifiers from the store.
Finally, $\mathtt{alive}_C$ is used to check if a value with a certain identifier is an element of the \emph{store}.

\begin{figure}[t]
    \begin{align*}
        \mathtt{to\_history}_C&: \Omega_r C \setprod (\tstr \setprod \flist \tnat) \longrightarrow \Omega_r C\\
        \mathtt{to\_history}_C&\left(\langle Q, S, H, I\rangle, rulename, i_1, ..., i_n\right) =\\
            &\quad\langle Q, S, H \cup \left\{\left(rulename, i_1, ..., i_n\right)\right\}, I\rangle
    \end{align*}
    \caption{Operations on the \emph{history} of a state}
    \label{fig:states:history}
\end{figure}

\begin{definition}[Operations on the \emph{history}]\label{fig:states:history}
    The function in \autoref{fig:states:history} define operations on the \emph{propagation history} of a state.
\end{definition}
The function $\mathtt{to\_history}_C$ adds a record to the propagation history in order to prevent the configuration represented by the record to fire again.

  \section{Instance with \emph{refined} semantics}\label{sec:instance}
In this section, we want to introduce the algorithms, that implement a FreeCHR instance with \emph{refined} operational semantics.
The algorithms are meant as blueprints for actual implementations, as well as a for theoretical considerations.
Implementations examples are available on \emph{GitLab}\footnote{\url{https://gitlab.com/freechr}}.

We begin with an abstract matching algorithm, derived from the \emph{refined} operational semantics of CHR.

\begin{definition}[Refined matching algorithm]\label{def:match}
    \autoref{alg:match} shows the abstract refined matching algorithm for FreeCHR.
    \begin{algorithm*}[t]
        \begin{algorithmic}[1]
            \Function{match}{$\mathit{name}$,$\mathit{head}$,$\mathit{guard}$,$i_a$,$c_a$,$\mathit{store}$,$\mathit{history}$}
                \State $\left[h_1, ..., h_n\right] \leftarrow head$
                \For{$j$ \textbf{from} $n$ \textbf{down to} $1$}
                    \label{alg:match:loop1}
                    \For{\textbf{each} $\left[\left(i_1, c_1\right), ..., \left(i_{j-1}, c_{j-1}\right), \left(i_{a}, c_{a}\right), \left(i_{j+1}, c_{j+1}\right), ..., \left(i_{n}, c_{n}\right)\right] \sqsubseteq \mathit{store}$}
                        \label{alg:match:loop2}
                        \If{$\neg \left(\mathit{h}_1\left(c_1\right) \wedge ... \wedge \mathit{h}_{j-1}\left(c_{j-1}\right) \wedge \mathit{h}_{j}\left(c_a\right) \wedge \mathit{h}_{j+1}\left(c_{j+1}\right) ...\wedge \mathit{h}_{n}\left(c_{n}\right) \right)$}
                            \label{alg:match:head}
                            \Continue
                        \EndIf
                        \If{$\left(\mathit{name}, i_1, ..., i_{j-1}, i_a, i_{j+1}, ..., i_n\right) \in \mathit{history}$}
                            \label{alg:match:history}
                            \Continue
                        \EndIf
                        \If{$\neg \mathit{guard}(c_1, ..., c_{j-1}, c_a, c_{j+1}, ..., c_n)$}
                            \label{alg:match:guard}
                            \Continue
                        \EndIf
                        \State \Return $\left[\left(i_1, c_1\right), ..., \left(i_{j-1}, c_{j-1}\right), \left(i_{a}, c_{a}\right), \left(i_{j+1}, c_{j+1}\right), ..., \left(i_{n}, c_{n}\right)\right]$
                        \label{alg:match:success}
                    \EndFor
                \EndFor
                \State \Return $\bot$
                \label{alg:match:failure}
            \EndFunction
        \end{algorithmic}
        \caption{Refined matching algorithm}
        \label{alg:match}
    \end{algorithm*}
\end{definition}

\autoref{alg:match} searches for an applicable configuration of values for a given rule.
The \emph{name}, \emph{head} and \emph{guard} of the rule, as well as the currently active value $(i_a, c_a)$, \emph{store} and \emph{history} of the current state are passed as arguments.
The main loop beginning in Line \ref{alg:match:loop1} implements the \textsc{default} transition for a single rule by iterating from right to left through the patterns of the head.
The inner loop beginning in Line \ref{alg:match:loop2} implements the search for an applicable configuration with $(i_a, c_a)$ matched to the pattern in the $j^{\text{th}}$ position.
The conditions in Lines \ref{alg:match:head} and \ref{alg:match:guard} essentially check, if the found matching constitutes a valid instance of the rule, as required by the \textsc{apply} transition.
Line \ref{alg:match:history} checks if this configuration already fired.
If every condition is met, the found configuration is returned.
If the main loop terminates, there was no valid matching.
In this case the next rule will be tried if there is one.

To actually implement \autoref{alg:match}, we used a depth-first-search with backtracking in our implementations, inspired by van Weert \cite{vanweert2010efficient}.

\begin{definition}[\textsc{rule} function]
    \autoref{alg:instance:rule} shows the execution algorithm for singular FreeCHR rules with refined semantics.

    \begin{algorithm}[t]
        \begin{algorithmic}[1]
            \Function{rule}{$\mathit{name}$, $\mathit{kept}$, $\mathit{removed}$, $\mathit{guard}$, $\mathit{body}$, $\mathit{is\_last?}$, $s$}
                \Loop
                    \label{alg:instance:rule:cleanuploop}
                    \If{$\mathtt{query}(s) \equiv []$}
                        \Return $\left(s, \bfalse\right)$
                        \label{alg:instance:rule:emptyqueue}
                    \EndIf
                    \State $(i_a, c_a) : \_ \leftarrow \mathtt{query}(s)$
                        \label{alg:instance:rule:topqueue}
                    \If{$i = \bot$}
                        $(i_a, s) \leftarrow \mathtt{activate}(s)$
                        \label{alg:instance:rule:activate}
                    \EndIf
                    \If{$\mathtt{alive}(s, i_a)$}
                        \Break
                        \label{alg:instance:rule:alive}
                    \EndIf
                    
                    \State $s \leftarrow \mathtt{pop\_query}(s)$
                        \label{alg:instance:rule:deadactive}
                \EndLoop
                    \label{alg:instance:rule:cleanuploop:end}
                \State $M \leftarrow \textsc{match}($
                    \label{alg:instance:rule:matching}
                    \State \quad $\mathit{name}, \mathit{kept} \diamond \mathit{removed}, \mathit{guard},$
                    \State \quad $i_a, c_a,$
                    \State \quad $\mathtt{store}(s), \mathtt{history}(s)$
                    \State $)$
                \If{$M \equiv \bot$}
                    \If{$\mathit{is\_last?}$} $s \leftarrow \mathtt{pop\_query}(s)$ \EndIf
                    \label{alg:instance:rule:drop}
                    \State \Return $(s, \bfalse)$ 
                    \label{alg:instance:rule:nomatching}
                \EndIf
                \State $\left[(i_1, c_1), ..., (i_n, c_n)\right] \leftarrow M$
                \State $s \leftarrow \mathtt{to\_history}(s, name, i_1, ..., i_n)$
                    \label{alg:instance:rule:recordpropagation}
                \State $s \leftarrow \mathtt{remove}(s, i_{\left|kept\right| + 1}, ..., i_n)$
                    \label{alg:instance:rule:remove}
                \State $s \leftarrow \mathtt{push\_query}(s, \mathit{body}(c_1, ..., c_n))$
                    \label{alg:instance:rule:query}
                \State \Return $(s, \btrue)$
                    \label{alg:instance:rule:return}
            \EndFunction
        \end{algorithmic}
        \caption{\textsc{rule} Algorithm}
        \label{alg:instance:rule}
    \end{algorithm}
\end{definition}

The \textsc{rule} function implements the application of a rule to the state.
The \textit{name}, \textit{kept} head, \textit{removed} head, \textit{guard} and \textit{body} of a rule, as well as a Boolean flag \textit{is\_last?} and the current state \textit{s} are passed as arguments.

The loop beginning in Line \ref{alg:instance:rule:cleanuploop} performs some preliminary checks and cleanups.
Line \ref{alg:instance:rule:emptyqueue} terminates the function if the query is empty, because in this case there is no transition defined by the refined operational semantics.
Line \ref{alg:instance:rule:activate} checks if the top value of the query is not yet activated and activates it, if so.
This essentially performs the \textsc{activate} transition.
Line \ref{alg:instance:rule:alive} checks if the value is actually alive.
If not, it is dropped from the query in Line \ref{alg:instance:rule:deadactive}, effectively performing the \textsc{drop} transition.
Line \ref{alg:instance:rule:matching} calls the \textsc{match} function.
Note that the concatenated head $\left(\mathit{kept} \diamond \mathit{removed}\right)$ is passed for the $\mathit{head}$ parameter.
If no valid matching was found, the function terminates.
If the flag \textit{is\_last?} is $\btrue$, this rule is assumed to be the last rule of the program.
A positive check in Line \ref{alg:instance:rule:drop} hence means that there are no matchings with $(i_a, c_a)$ as the active value.
The value is hence dropped from the query, which effectively performs the \textsc{drop} transition.
Finally, Lines \ref{alg:instance:rule:recordpropagation} to \ref{alg:instance:rule:query} perform the operations defined by \textsc{apply}.
First the applied configuration is recorded in the history.
Then the values matched to the removed head are removed from the store.
Last, the values generated by the body are queried and the successor state is returned.
Note, that the function also returns a Boolean flag which indicates if a rule was actually applied ($\btrue$) or not ($\bfalse$).

\begin{algorithm}[t]
    \begin{algorithmic}[1]
        \Function{compose}{$p_1, ..., p_n, \mathit{is\_last?}, s$}
            \For{$i$ \textbf{from} $1$ \textbf{to} $n$}
                    \label{alg:instance:compose:mainloop}
                \State $(s, \mathit{rule\_applied?}) \leftarrow p_i(i = n \wedge \mathit{is\_last?}, s)$
                    \label{alg:instance:compose:apply}
                \If{$\mathit{rule\_applied?}$} 
                    \Return $(s, \btrue)$
                        \label{alg:instance:compose:success}
                \EndIf
                    \label{alg:instance:compose:applycheck}
            \EndFor
            \State \Return $(s, \bfalse)$
                \label{alg:instance:compose:failure}
        \EndFunction
    \end{algorithmic}
    \caption{\textsc{compose} Algorithm}
    \label{alg:instance:compose}
\end{algorithm}

\begin{definition}[\textsc{compose} function]
    \autoref{alg:instance:compose} shows the execution algorithm to compose multiple FreeCHR programs with refined semantics.
\end{definition}
The \textsc{compose} function implements the traversal of a program from top to bottom.
The component program $p_1, ..., p_n$, as well as a Boolean flag $is\_last?$ and the current state, is passed as an argument.
The loop iterates $i$ over the indices of the passed programs.
This, together with the \textsc{match} function called by \textsc{rule} implements a full iteration of the pattern index of the currently active value, as performed by \textsc{default}.
Line \ref{alg:instance:compose:apply} tries to apply $p_i$ to the state.
If this succeeds, the successor state is returned in Line \ref{alg:instance:compose:success}.
Otherwise, the loop continues.
The Boolean flag \textit{is\_last?} is used to propagate the information, which rule is the last of the program, to the rightmost (\IE last) component $p_n$ of the composite $p_1 \chrcomp ... \chrcomp p_n$.
This is done by passing the value of $(i = n \wedge \mathit{is\_last?})$ to $p_i$.
If the loop terminates, no applicable program was found and the state is returned with the respective flag.
Note, that this implementation deviates from the original $\omega_r$, in that it behaves as if the pattern index of the active value was reset to $1$, once a rule was applied.
This has no significant consequences on semantics since repeated applications of the same configuration are already prevented by the propagation history and applications with subsequently added values will be performed anyway when they become active.
It, however, has a severe impact on performance which we, for the sake of simplicity, will ignore for now and address in future work.

\begin{algorithm}[t]
    \begin{algorithmic}[1]
        \Function{run}{$p, s$}
            \While{$\mathtt{query}(s) \not\equiv \left[\right]$}
                \State $(s, \_) \leftarrow p(s)$
                    \label{alg:instance:run:apply}
            \EndWhile
            \State \Return $s$
                \label{alg:instance:run:return}
        \EndFunction
    \end{algorithmic}
    \caption{\textsc{run} Algorithm}
    \label{alg:instance:run}
\end{algorithm}

\begin{definition}[\textsc{run} driver function]
    \autoref{alg:instance:run} shows the driver function for FreeCHR programs.
\end{definition}
Finally, the \textsc{run} function applies the program $p$ to the state $s$, until the query is empty.
Since $p$ is the root of the program, it gets $\btrue$ passed as its $\mathit{is\_last?}$ flag.

\begin{figure}[t]
    \begin{center}
        \small
        \begin{align*}
                                                                &\ \langle [(\bot, 6), (\bot, 9)], \emptyset, \emptyset, 1 \rangle\\
            \xmapsto{\textsc{rule}(\textit{"zero"}, ...)}       &\ \langle [(1, 6), (\bot, 9)], \left\{(1, 6)\right\}, \emptyset, 1 \rangle\\
            \xmapsto{\textsc{rule}(\textit{"subtract"}, ...)}   &\ \langle [(\bot, 9)], \left\{(1, 6)\right\}, \emptyset, 2 \rangle\\
            \xmapsto{\textsc{rule}(\textit{"zero"}, ...)}       &\ \langle [(2, 9)], \left\{(1, 6), (2, 9)\right\}, \emptyset, 3 \rangle \\
            \xmapsto{\textsc{rule}(\textit{"subtract"}, ...)}   &\ \langle [(\bot, 3), (2, 9)], \left\{(1, 6)\right\}, \left\{\textcolor{red!70!black}{("subtract", 1, 2)}\right\}, 3 \rangle \\
            \xmapsto{\textsc{rule}(\textit{"zero"}, ...)}       &\ \langle [(3, 3), (2, 9)], \left\{(1, 6), (3, 3)\right\}, \left\{...\right\}, 4 \rangle \\
            \xmapsto{\textsc{rule}(\textit{"subtract"}, ...)}   &\ \langle [(\bot, 3), (3, 3), (2, 9)], \left\{(3, 3)\right\}, \left\{..., \textcolor{red!70!black}{("subtract", 3, 1)}\right\}, 4 \rangle \\
            \xmapsto{\textsc{rule}(\textit{"zero"}, ...)}       &\ \langle [(4, 3), (3, 3), (2, 9)], \left\{(3, 3), (4, 3)\right\}, \left\{...\right\}, 5 \rangle \\
            \xmapsto{\textsc{rule}(\textit{"subtract"}, ...)}   &\ \langle [(\bot, 0), (4, 3), (3, 3), ...], \left\{(3, 3)\right\}, \left\{..., \textcolor{red!70!black}{("subtract", 3, 4)}\right\}, 5 \rangle \\
            \xmapsto{\textsc{rule}(\textit{"zero"}, ...)}       &\ \langle [(5, 0), (4, 3), (3, 3), ...], \left\{(3, 3)\right\}, \left\{..., \textcolor{red!70!black}{("zero", 5)}\right\}, 6 \rangle \\
            \xmapsto{\textsc{rule}(\textit{"zero"}, ...)}       &\ \langle [(3, 3), (2, 9)], \left\{(3, 3)\right\}, \left\{...\right\}, 6 \rangle \\
            \xmapsto{\textsc{rule}(\textit{"subtract"}, ...)}   &\ \langle [(2, 9)], \left\{(3, 3)\right\}, \left\{...\right\}, 6 \rangle \\
            \xmapsto{\textsc{rule}(\textit{"zero"}, ...)}       &\ \langle [], \left\{(3, 3)\right\}, \left\{...\right\}, 6 \rangle \\
            \xmapsto{\textsc{rule}(\textit{"subtract"}, ...)}   &\ \langle [], \left\{(3, 3)\right\}, \left\{...\right\}, 6 \rangle
        \end{align*}
    \end{center}
    \caption{Execution of the Euclidean algorithm implemented in FreeCHR with initial query $[6,9]$}
    \label{fig:freechr:gcd:exec}
\end{figure}

\begin{example}[Greatest common divisor in FreeCHR, executed]\label{ex:freechr:gcd:exec}
    We want to repeat \autoref{ex:chr:gcd:exec} with the defined FreeCHR instance and the program defined in \autoref{ex:freechr:gcd}.
    Again, we use $[6,9]$ as our initial query.
    If we call
    \begin{align*}
        \textsc{run}(\textsc{gcd}, \langle [6, 9], \emptyset, \emptyset, 1 \rangle)
    \end{align*}
    with $\textsc{gcd} = \kata{exec_r}(\mathit{zero} \chrcomp \mathit{substract})$, \IE
    \begin{align*}
        \textsc{gcd} = \textsc{compose}(&\textsc{rule}(\textit{"zero"}, \eps, \left[\lambda n. n = 0\right], (\lambda n. \btrue), (\lambda n. \eps)), \\
            &\textsc{rule}(\textit{"subtract"}, \left[\lambda n. 0 < n\right], \left[\lambda m. 0 < m\right],\\
            &\quad (\lambda n\ m. n \leq m), (\lambda n\ m. \left[m-n\right])))
    \end{align*}
    we get the transitions shown in \autoref{fig:freechr:gcd:exec}.
 
    First, the \emph{zero}-rule activates the value $6$.
    As it is not able to find a matching it terminates, and the \textsc{compose} function tries to apply \emph{subtract} to it.
    In this call, the loop in the beginning \autoref{alg:instance:rule} terminates, as the top queried value is already active.
    Since \emph{subtract} is the last rule of the program and can not fire either, it drops the active value from the query.
    Now, $9$ is activated by \emph{zero}, but the rule can not fire, and \textsc{compose} hands the state over to \emph{subtract}.
    This rule is able to fire and the call of \emph{subtract} removes $(2, 9)$ and queries $3$. 
    Again, \emph{zero} activates the top queried value, terminates because it is not equal to $0$, and hands over to \emph{subtract} which removes $(1,6)$ and queries $3$.
    Another time, \emph{zero} activates the top value and terminates.
    \emph{subtract} fires, removes the newly added $(4,3)$ and queries $0$.
    Now, \emph{zero} activates the $0$ on top of the query and fires, instantly removing $(5,0)$ from the store.
    At this point, \emph{zero} cleans up all values until $(3,3)$, as it is still alive, but is unable to fire, as is \emph{subtract}.
    It hence drops this value from the query.
    Finally, \emph{zero} cleans up the query completely and terminates, because the query is empty.
    \emph{subtract} then encounters an empty query as well and terminates, as does \textsc{compose} and ultimately \textsc{run}.
\end{example}
  \section{Proof of correctness}\label{sec:soundness}
In this section, we want to prove that the FreeCHR instance defined by \autoref{alg:instance:rule} and \autoref{alg:instance:compose} is a correct concretization of the \emph{very abstract} operational semantics $\omstara$.
We do this by proving that every transition performed by the instances can be performed by $\omstara$.

We first want to define the transition system that is implied by the FreeCHR instance 
\begin{align*}
    \langle \left(\Omega_r C, \tbool\right)^{\tbool \times \Omega_r C}, \left[\textsc{rule}, \textsc{compose}\right] \rangle
\end{align*}

\begin{definition}[Instance transition system]\label{def:refined:transitions}
    With
    \begin{align*}
        exec_r = \left[\textsc{rule}, \textsc{compose}\right]
    \end{align*}
    \autoref{alg:instance:rule} and \autoref{alg:instance:compose} define the labelled transition system
    \begin{align*}
        \omstarr = \langle \Omega_r C, \mu\chr_C, (\freetrr{})^{*} \rangle
    \end{align*}
    where the relation $(\freetrr{})$ is defined as
    \begin{align*}
        (\freetrr{}) = \left\{ s \freetrr{p} s'\ :\ \kata{exec_r}(p)(\btrue, s) \equiv (s',\_) \right\}
    \end{align*}
    and $(\freetrr{})^{*}$ is the reflexive-transitive closure of $(\freetrr{})$.
\end{definition}
According to \autoref{def:refined:transitions} every transition $s \freetrr{p} s'$ is valid, \IFF applying the catamorphism $\kata{exec_r}$ to a program $p \in \chr_C$ results in a function that maps $s$ to $s'$.\footnote{We discard the Boolean flag added to $s'$ by \textsc{rule} and \textsc{compose}, as it is irrelevant outside these functions.}
We set the Boolean flag to $\btrue$, since we assume $p$ to be the whole (root) program, as called by \textsc{run}.

\subsection{Refinement}
Since the $\omstara$ and $\omstarr$ operate on different kinds of states, we need a function abstracting states of $\Omega_r C$ to $\fmultiset C$.
\begin{definition}[Abstraction function]\label{def:abstract}
    We call 
    \begin{align*}
        \abstractr &: \Omega_r C \longrightarrow \fmultiset C\\
        \abstractr(s) &= \left\{ c : (\_, c) \in \mathtt{store}(s) \right\} \msetplus \left\{ c : \left(\bot, c\right) \in \mathtt{query}(s) \right\}
    \end{align*}
    the \emph{abstraction function} of $\Omega_r$ to $\fmultiset$. 
\end{definition}

The function $\abstractr$ takes all inactive values from the query and all values from the store without decoration and constructs a multiset from them.
We now state and prove our main theorem, which connects the new definitions proposed in this work back to our established definitions.
\begin{theorem}\label{theorem:completeness}
    $\omstara$ is $\abstractr$-sound \WRT $\omstarr$.
\end{theorem}
\begin{proof}
    We prove that for any $s, s' \in \Omega_r C$ and $p \in \mu\chr_C$
    \begin{align*}
        s \freetrr{p} s' \in (\freetrr{})^{*} \Longrightarrow \abstractr(s) \freetr{p} \abstractr(s') \in (\freetr{})^{*}
    \end{align*}
    by first showing that
    \begin{align*}
        \kata{exec_r}(p)(\mathit{lst?}, s) \equiv s' \Longrightarrow \abstractr(s) \freetr{p} \abstractr(s') \in (\freetr{})^{*}
    \end{align*}
    for $\mathit{lst?} \in \tbool$, via induction over the structure of $p$.
    The rest follows from the properties of the reflexive-transitive closure.
    \begin{indbase}[$p = rule(N, k, r, g, b)$]
        We show that
        \begin{align*}
            &\textsc{rule}(N, k, r, g, b, \mathit{lst?}, s) \equiv s'\\
            &\quad \Longrightarrow \abstractr(s) \freetr{rule(N, k, r, g, b)} \abstractr(s') 
        \end{align*}

        \begin{case*}[return in Line \ref{alg:instance:rule:emptyqueue}]
            We differentiate between two cases: either the loop did not iterate on execution of the \Return statement, or it did.
            In the first case $s$ was left unchanged, hence $\textsc{rule}(N, k, r, g, b, s) \equiv s$.
            In the latter case, the only change to the state might have occurred in Line \ref{alg:instance:rule:deadactive}, since Line \ref{alg:instance:rule:activate} would guarantee $i_a$ to refer to a live value, causing Line \ref{alg:instance:rule:alive} to break the loop, after which Line \ref{alg:instance:rule:emptyqueue} will not be executed.

            Since we reached Line \ref{alg:instance:rule:deadactive}, $i_a$ refers to an already deleted value and given \autoref{def:abstract}, which only considers inactive values (those with $\bot$ as their identifier) and values in the store, we can conclude that, although $s$ was altered, $\abstractr(\textsc{rule}(N, k, r, g, b, s)) \equiv \abstractr(s)$.

            \begin{align*}
                    \abstractr(s) \freetr{rule(N, k, r, g, b)} \abstractr(s')
            \end{align*}
            is hence a reflexive element in $(\freetr{})^{*}$.
            \hfill\checkmark
        \end{case*}

        \begin{case*}[return in Line \ref{alg:instance:rule:nomatching}]
            Line \ref{alg:instance:rule:activate} and \ref{alg:instance:rule:deadactive} were already discussed above.
            The changes performed by Line \ref{alg:instance:rule:drop} do also not alter the result of $\abstractr$, independent of the value of \textit{lst?}.

            \begin{align*}
                \abstractr(s) \freetr{rule(N, k, r, g, b)} \abstractr(s')
            \end{align*}
            is hence a reflexive element in $(\freetr{})^{*}$.
            \hfill\checkmark
        \end{case*}

        \begin{case*}[return in Line \ref{alg:instance:rule:return}]
            For this final case we must additionally consider Lines \ref{alg:instance:rule:recordpropagation}, \ref{alg:instance:rule:remove} and \ref{alg:instance:rule:query}.
            Line \ref{alg:instance:rule:recordpropagation} can be discarded, as it only modifies the history which is ignored by $\abstractr$.
            By Lines \ref{alg:match:head} and \ref{alg:match:guard} of \autoref{alg:match}, we can assume that
            \begin{align*}
                &k_1(c_1) \wedge ... \wedge k_{n}(c_{n+m})\tag{Kept head}\\
                \wedge\ &r_1(c_{n+1}) \wedge ... \wedge r_{m}(c_{n+m})\tag{Removed head}\\
                \wedge\ &g(c_1, ..., c_{n+m})\tag{guard}\\
                \equiv\ &\btrue
            \end{align*}
            By Line \ref{alg:match:loop2} of \autoref{alg:match} we can also assume that $(i_a, c_a) \in M$.
            Line \ref{alg:instance:rule:remove} will remove any value with identifiers $i_{n+1}$ to $i_{n+m}$.
            Hence, by passing the statement we will have modification
            \begin{align*}
                \abstractr(s) \xmapsto{Line \ref{alg:instance:rule:remove}} \abstractr(s) \setminus \left\{c_{n+1}, ..., c_{n+m}\right\}
            \end{align*}
            Line \ref{alg:instance:rule:query} will then add the values created by $b\left(c_1, ..., c_{n+m}\right)$ to the query as inactive values.
            Hence, by passing this statement we will have modification
            \begin{align*}
                &\abstractr(s) \setminus \left\{c_{n+1}, ..., c_{n+m}\right\}\\
                \quad \xmapsto{\text{Line \ref{alg:instance:rule:query}}}\ 
                    &\abstractr(s) \setminus \left\{c_{n+1}, ..., c_{n+m}\right\} \msetplus b\left(c_1, ..., c_{n+m}\right)\\
            \end{align*}
            By Line \ref{alg:match:loop2} of \autoref{alg:match} we can assume that
            \begin{align*}
                \left\{c_1, ..., c_{n+m}\right\} \subseteq \abstractr(s)
            \end{align*}
            We can hence rewrite $\abstractr(s)$ as $\left\{c_1, ..., c_{n+m}\right\} \msetplus \Delta s$ and
            \begin{align*}
                \abstractr(s) \setminus \left\{c_{n+1}, ..., c_{n+m}\right\} \msetplus b\left(c_1, ..., c_{n+m}\right)
            \end{align*}
            as
            \begin{align*}
                \left\{c_1, ..., c_{n}\right\} \msetplus b\left(c_1, ..., c_{n+m}\right) \msetplus \Delta s
            \end{align*}
            Thus, we can provide a proof
            \begin{align*}
                \begin{prooftree}[small]
                    \hypo{k_1\left(c_1\right) \wedge ... \wedge k_{n}\left(c_{n}\right) \wedge r_1\left(c_{n+1}\right) \wedge ... \wedge r_{m}\left(c_{n+m}\right) \wedge g\left(c_1, ..., c_{n+m}\right) \equiv \btrue}
                    \infer1[apply]{\left\{c_1, ..., c_{n+m}\right\} \msetplus \Delta s
                    \freetr{rule(N, k, r, g, b)}
                    \left\{c_1, ..., c_{n}\right\} \msetplus b\left(c_1, ..., c_{n+m}\right) \msetplus \Delta s}
                \end{prooftree}
            \end{align*}
            \hfill\checkmark
        \end{case*}
    \end{indbase}

    \begin{indstep}[$p = p_1 \chrcomp ...\chrcomp p_n$]
        As induction hypothesis we assume for all $i \in \left\{1, ..., n\right\}$, $\mathit{lst?} \in \tbool$ and any $s, s' \in \Omega_r C$ that
        \begin{align}
            \kata{exec_r}(p_i)(\mathit{lst?}, s) \equiv s'
            \Longrightarrow 
            \abstractr(s) \freetr{p_i} \abstractr(s')
            \label{eq:completeness:noTransitiveSteps:indhypo}
        \end{align}
        We show that
        \begin{align*}
            &\textsc{compose}(\kata{exec_r}(p_1), ..., \kata{exec_r}(p_n), \mathit{lst?}, s) \equiv s'\\
            &\quad \Longrightarrow \abstractr(s) \freetr{p_1 \chrcomp ...\chrcomp p_n} \abstractr(s')
        \end{align*}

        \begin{case*}[return in Line \ref{alg:instance:compose:failure}]
            If $s'$ was returned in Line \ref{alg:instance:compose:failure}, the only modifying statements were executed in Lines \ref{alg:instance:rule:emptyqueue}, \ref{alg:instance:rule:activate}, \ref{alg:instance:rule:deadactive} and \ref{alg:instance:rule:drop}.
            We already discussed above that they do not affect the result of $\abstractr$.

            \begin{align*}
                \abstractr(s) \freetr{p_1 \chrcomp ...\chrcomp p_n} \abstractr(s')
            \end{align*}
            is hence a reflexive element in $(\freetr{})^{*}$.
            \hfill\checkmark
        \end{case*}

        \begin{case*}[return in Line \ref{alg:instance:compose:success}]
            If $s'$ was returned in Line \ref{alg:instance:compose:success}, it is the result of applying a subprogram $p_i$ to $s$.
            With the induction hypothesis (\ref{eq:completeness:noTransitiveSteps:indhypo}) we can construct a proof
            \begin{align*}
                \begin{prooftree}[small]
                    \hypo{\abstractr(s) \xmapsto{p_i} \abstractr(s')}
                    \infer1[step]{\abstractr(s) \freetr{p} \abstractr(s')}
                \end{prooftree}
            \end{align*}
        \end{case*}
    \end{indstep}
\end{proof}

\autoref{theorem:completeness} establishes our algorithmic representation of $\omstarr$ as a valid concretization of $\omstara$,
and thereby implementations according the definitions in \autoref{sec:instance} as correct implementations of FreeCHR and in consequence of CHR.
  \section{Related work}\label{sec:related-work}
\subsection{Constraint Handling Rules}
\subsubsection{Embeddings}
The first approach to embed CHR into a host language was via source-to-source transformation.
Holzbaur \EA \cite{holzbaur1998compiling,holzbaur2000prolog} and Schrijvers \EA \cite{schrijvers2004leuven}, \EG translate CHR via Prolog's macro system.
Similarily, Abdennadher \EA \cite{abdennadher2002jack} and van Weert \EA \cite{vanweert2005leuven} use a precompiler to translate CHR programs into Java,
Wuille \EA \cite{wuille2007cchr} into C, Nogatz \EA \cite{nogatz2018chr} into Javascript, and Barichard \cite{barichard2024chr} into C++.

van Weert \cite{vanweert2008chr} introduces compilation schemes for imperative languages, upon which, \EG Nogatz \EA \cite{nogatz2018chr} builds.
The work of van Weert \cite{vanweert2010efficient} also provides compilation schemes for imperative languages, but optimizes the matching process by performing it lazily.
The algorithms do not first compute a full matching and check it against the patterns and the guard but collect it successively, and partially check the guard if possible.

The inherent disadvantage of an approach using source-to-source compilation is the need for a precompiler in the build chain if the host language does not have a sufficiently expressive macro system like Prolog or LISP.
It comes with the cost of more sources of errors and an additional dependency that is often rather tedious to fulfill.
Hanus \EA \cite{hanus2015chr} approaches this problem by extending the Curry compiler to be able to compile CHR code.
The obvious problem with this approach is that it is in most cases not viable or possible to extend the compiler of the host language.

A relatively new approach by Ivanović \cite{ivanovic2013implementing} was to implement CHR as an internal language in Java.
This has one major advantage: CHR can now simply be imported as a library, similarly as if implemented by a macro system.
Wibiral \cite{wibiral2022javachr} further builds upon this idea and introduces the idea of explicitly composing CHR programs of singular rules by an abstract and modular execution strategy and describing rules through anonymous functions.
This was our main inspiration and FreeCHR aims to improve and generalize this idea.

\subsubsection{Operational semantics}
Duck \EA \cite{duck2004refined} first formalized the behavior of existing implementations which were mostly derived from but not exactly true to existing formal definitions of operational semantics.
The techniques used in the implementations were also generalized and improved upon by van Weert \EA \cite{vanweert2008chr,vanweert2010efficient}.
Duck \cite{duck2005compilation} standardized call-based semantics for CHR, especially in logical programming languages.

\subsection{Algebraic language embeddings}
The idea of algebraic embeddings of a domain specific language (DSL) into a host language was first introduced by Hudak \cite{hudak1998modular}.
The style in which the languages are embedded was later called \emph{tagless}, as it does not use an algebraic data type, to construct an abstract syntax tree (see, \EG Carette \EA \cite{carette2007finally}).
The \emph{tags} are the constructors of the data type which defines the syntax of the language.
Instead, the embedding directly defines the functions which implement the semantics of the language.
This is, defining the free $F$-algebra versus defining the concrete $F$-algebras, for a functor $F$ which defines the syntax of the language.
The advantage of embedding a DSL in this way is that it does not rely on external build tools, but can instead be easily implemented and used as a library.
It also enables the use of any features the host language offers, without any additional work.
With an external source-to-source compiler, it would be necessary to re-implement at least the syntax of any desired host language features.

Hofer \EA \cite{hofer2008polymorphic,hofer2010modular} then extended this idea by using type families in order to provide more flexibility concerning the semantics of the language.

\subsection{Logic and constraint based languages and formalisms}
CHR was initially designed as a tool to implement constraint solvers for user-defined constraints.
Hence, its domain intersects with those of \emph{Answer Set Programming} (ASP) and \emph{Constraint Logic Programming} (CLP).
However, CHR can rather be understood as a tool in combination with ASP or especially CLP, as it provides an efficient language to implement constraint solvers, than an alternative approach.
On the other hand, CHR has already exceeded its original purpose and developed more towards a general purpose language.
It hence also exceeded the domain of ASP and CLP.

Another relevant logic based language is \emph{miniKanren} \cite{byrd2009relational}.
\emph{miniKanren} is a family of EDSLs for relational and logic programming.
There exists a myriad of implementations for plenty of different host languages as well as formal descriptions of operational semantics (see \EG Rozplokhas \EA \cite{rozplokhas2020certified}).
The \emph{miniKanren} language family seems to suffer from similar issues as traditional \emph{Constraint Handling Rules} does.
As there is no unifying embedding scheme, there is an inherent disconnect between any implementation and the formally defined semantics.
Applying an approach similar to ours might be beneficial to the \emph{miniKanren} project as well, especially as it generally is implemented in a way similar to the methods described by Hudak \cite{hudak1998modular} or Carette \EA \cite{carette2007finally}.

  \section{Limitations}\label{sec:limitations}
In this section, we want to briefly discuss some limitations of our approach.

\subsection{Logical variables}
First, as described in earlier work \cite{rechenberger2023freechr}, FreeCHR is based on the \emph{positive}, \emph{range restricted} and \emph{ground} segment of CHR.
This subset of CHR is commonly used as the target for embeddings of other (rule-based) formalisms into CHR \cite{fruehwirth2025principles}.
\emph{Positive} effectively means that the body of the rule does not cause any side effects and especially guarantees that computations do not fail.
\emph{Range-restricted} means that instantiating all variables of the head will ground the whole rule.
This also maintains the \emph{groundness} of the segment of CHR which requires that the input and output of a program are ground.

CHR was initially implemented and formalized with Prolog in mind.
In Prolog, logical variables can be considered a native data type.
Hence, by abstracting from host languages, we viewed logical variables as a possible, but not guaranteed feature, which we treated like any other data type.
We thus assume FreeCHR rules to be \emph{ground} in a logical sense.
From a practical angle, if the host language provides logical variables, they can be freely used within a FreeCHR instance.

\subsection{Performance}
In order to simplify formal considerations, we kept the provided execution algorithms as close to the definition of the \emph{refined} operational semantics and as simple as possible.
We hence applied no optimization whatsoever, be it very simple or more complex ones.

Major bottlenecks to performance in CHR are matching and the propagation history \cite{vanweert2010efficient}.
Depending on the way CHR is implemented, it is possible to deconstruct parts of the program and, for example, check only relevant parts of the guard of a rule upon matching.
Since FreeCHR models guard and patterns as functions, it is harder to access parts of it, without more advanced programming techniques like meta-programming.
Easier to implement are optimization on the housekeeping of the propagation history.
A very simple optimization would be to only add or check for a record, if the firing rule is a propagation rule.
Other more advanced techniques could be applied to remove records which could not fire again, because the respective values were already removed.
Both ideas help to keep the time spent searching the history as short as possible.
By using iterator-based solutions for matching, we can also omit the propagation history all together \cite{vanweert2010efficient}.
The idea is to (lazily) generate all possible configurations and execute them until the respective active value is removed.

Optimizations on the execution algorithms and proofs of their correctness \WRT the baseline presented here are topic of future work, as are benchmarks to analyze the effect of the optimization and the performance \WRT existing CHR implementations.

\subsection{Abstract matching algorithm}
For the same reason for which we did not include any optimizations, we included only the very abstract matching algorithm in \autoref{alg:match}.
The presented algorithm serves as an easy to grasp baseline for formal considerations.
Another reason for this is that, depending on host language and preferred programming style, the concrete implementation of the matching algorithm may vary.
As mentioned above, we used a depth-first-search to implement the algorithm in our implementations in Haskell and Python.
We plan to formalize and analyze the actually implemented algorithm and prove its correctness \WRT \autoref{alg:match} in future work.

\section{Future Work}\label{sec:future_work}
Ongoing work is mainly concerned with formalizing the refined operational semantics $\omega_r^{\star}$ of FreeCHR structurally (\IE via inference rules) and proving soundness and completeness \WRT both, the algorithmic definition presented herein and the original definition of $\omega_r$.
Thereby, we prove that the algorithmic definition is a valid representation of the original $\omega_r$ and that the translation $\omega_r^{\star}$ is a valid concretization of $\omstara$.

Since FreeCHR is, in its core, driven by practical intentions, future work will be concerned with optimizing the algorithms presented herein and developing them to be competitive with existing CHR implementations, while at the same time, keep them formal rigorosity.
This, on the one hand, includes benchmarks to validate the effectiveness of optimizations, and on the other hand proofs to verify correctness \WRT the baseline presented herein.

\section{Conclusion}\label{sec:conclusion}
In this paper, we introduced an execution algorithm for FreeCHR which we derived from the \emph{refined} operational semantics of CHR.
We also established the presented algorithm as a valid concretization of the very abstract operational semantics of FreeCHR.

The presented algorithm serves a twofold purpose: first and foremost it provides a blueprint for implementations of FreeCHR, that are as expressive as existing implementations of CHR.
Second, it also provides an algorithmic description of the \emph{refined} operational semantics for FreeCHR and serves hence as a means for formal considerations.

Concluding, we presented in this work, to our knowledge, the first formalization of a fully expressive CHR embedding for which there exist formal proofs of correctness.

  \bibliographystyle{plain}
  \bibliography{local/bibliography}
\end{document}